\documentclass[aip,apl,reprint,amsmath,superscriptaddress]{revtex4-1}
\usepackage{graphicx}
\usepackage{amsmath,amsthm,amssymb,latexsym,amsfonts,times,mathrsfs,verbatim, lipsum}
\usepackage{bm}
\usepackage{color}

\begin{document}
\title{Broadband optomechanical transduction of nanomagnetic spin modes} 

\author{P.H. Kim}
\author{F. Fani Sani}
\author{M.R. Freeman}
\author{J.P. Davis}\email{jdavis@ualberta.ca}
\address{Department of Physics, University of Alberta, Edmonton, Alberta, Canada T6G 2E9}


\begin{abstract}  
The stable vortex state that occurs in micron-scale magnetic disks is one of the most interesting and potentially useful phenomenon in nanomagnetism.  A variety of tools have been applied to study the vortex state, and collective spin excitations corresponding to harmonic motion of the vortex, but to-date these tools have measured either strongly driven vortex resonances or have been unable to simultaneously measure static properties such as the magnetization.  Here we show that by combining the sensitivity of cavity optomechanics with the technique of torque mixing resonance spectroscopy, we are able to measure the magnetization, in-plane susceptibility, and spin resonances of individual vortices in the low-drive limit. These measurements elucidate the complex behavior of the vortex as it moves through the pinning landscape of the disk.  Furthermore, we observe gyrotropic resonances as high as 1.1 GHz, suggesting the use of engineering defects for applications such as microwave-to-optical wavelength conversion. 

\end{abstract}

\maketitle

Magnetometry enabled by micro and nanomechanical resonators has proven to be one of the most useful probes of nanomagnetism.  Torsional resonators have allowed the observation of such phenomenon as real-time creation and annihilation of single magnetic vortices,\cite{Davis10}  Barkhausen noise from a single defect,\cite{Burgess13} mechanical ferromagnetic resonance spectroscopy,\cite{Klein08} and collective spin modes using the technique of torque mixing resonance spectroscopy (TMRS).\cite{Los15}  Adding to this the sensitivity of cavity optomechanics, which allows thermomechanical torsional motion to be observed from room temperature\cite{Kim13,Wu17} down to millikelvin temperatures,\cite{Kim16} has the potential to uncover new physics, particularly in the stochastic domain, of nanomagnetism.  Here we show that by using a cavity optomechanical torque magnetometer\cite{Kim13} optimized for torque mixing resonance spectroscopy, we are able to observe spin modes in magnetically-soft permalloy disks above 1.1 GHz.  By comparing with micromagnetic simulations and direct-torque susceptometry measurements,\cite{Wu17} we show that these modes are consistent with the gyrotropic mode\cite{Park03,Novosad05,Perzlmaier05,Vogt11,Riley15} of the vortex sampling the disorder potential of the polycrystalline magnetic disk.\cite{Uhlig05,Burgess13}   Importantly, we are able to track these spin modes at all frequencies, that is we see no drop-outs in the spin resonances, which allows us to follow the behavior of the vortex as it traverses the disk.  In future studies, it should be possible to work backwards from these spin modes to recover the energy landscape of the magnetic structure. \cite{Burgess13}  We also note that this work amounts to conversion of UHF (ultra-high frequency) signals through spin resonances to telecom wavelengths.  Such wavelength conversion is a key topic in quantum technology applications,\cite{Bochmann13,Andrews14,Hisatomi16} and improvements to the existing device architecture may provide a promising route to high-efficiency wavelength conversion\cite{Hil12} through quadruply-resonant (optical, mechanical, spin, and microwave) devices.

\begin{figure}[b]
\centerline{\includegraphics[width=3.3in]{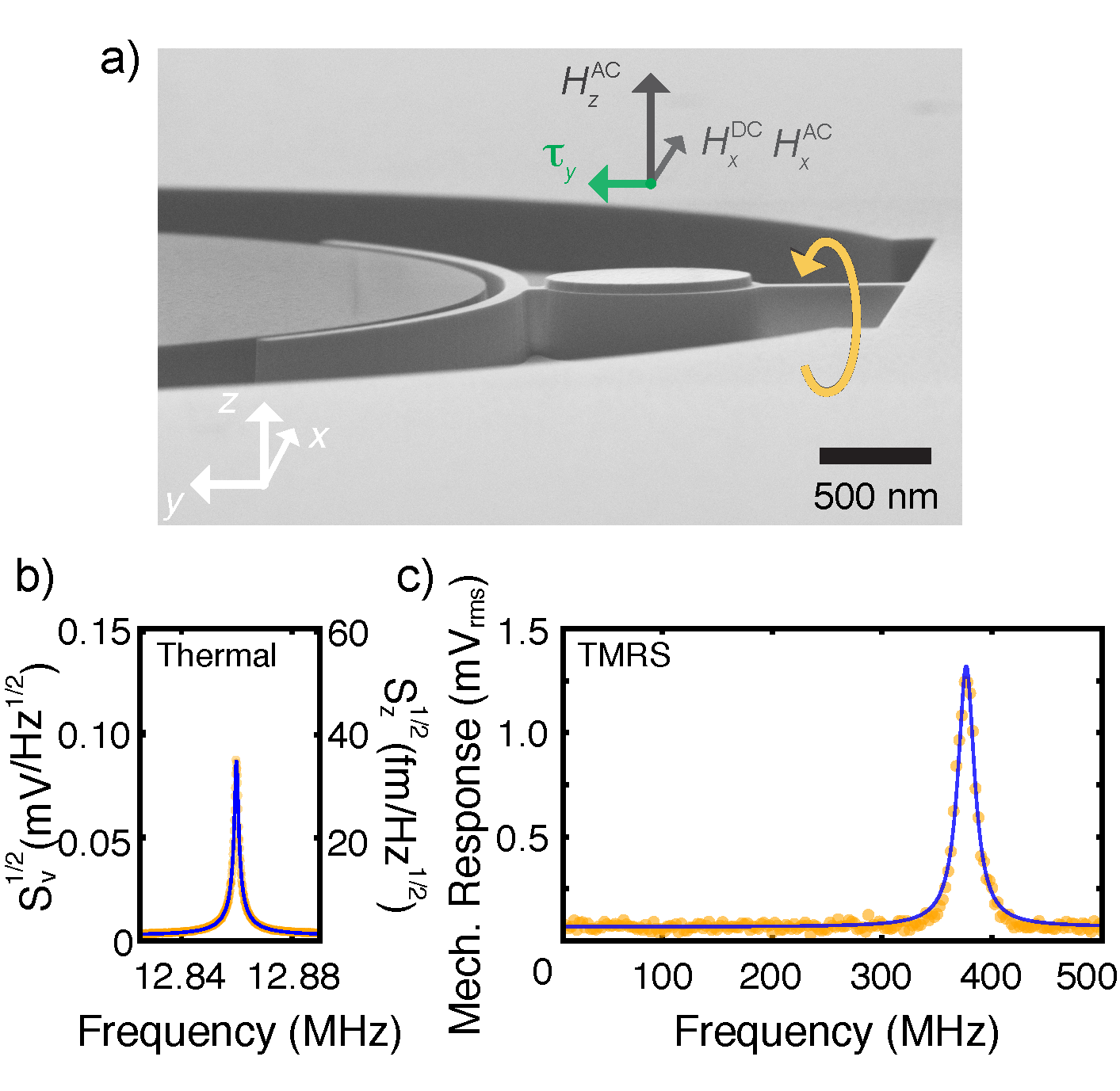}}
\caption{{\label{fig1}}Tilted SEM image of a 1.1 $\mu$m diameter permalloy disk fabricated onto a freestanding silicon optomechanical torsional resonator.  Optomechanical detection is read out through a dimpled tapered fiber\cite{Michael07,Hau14} that monitors an optical resonance of the whispering gallery mode microdisk, seen at the left.  AC magnetic fields can be applied to the permalloy disk in both the $x$ and $z$ directions, as well as a DC magnetic field in the plane of the disk.  These allow monitoring of both the direct torque, encoding the disks magnetization and in-plane susceptibility, as well as the spin mode resonances through torque mixing resonance spectroscopy.\cite{Los15}  b) The calibrated thermomechanical motion\cite{Hauer13} of the torsional mode at 12.8 MHz, and c) gyrotropic mode of the vortex detected via torque mixing resonance spectroscopy.}
\end{figure}

We present data from two devices, fabricated simultaneously on the same 500 $\mu$m thick silicon-on-insulator chip, with different permalloy disk diameters.  Nanofabrication is performed using two e-beam lithography steps.  The first defines the optical disks and mechanical resonators, followed by reactive ion etching to transfer this pattern to the single crystal silicon.  After etching, the resist is removed and a second e-beam lithography step is used to define a disk on the ``landing pad'' of the torsional resonator.  Up to this point fabrication is similar our previous torsional devices,\cite{Kim16,Kim17} but in this work we slowly deposit 50 nm of permalloy (Ni$_{80}$Fe$_{20}$) in ultra-high vacuum to produce low-defect density nanomagnetic disks.\cite{Davis10,Burgess13}  After permalloy deposition we perform lift-off, followed by HF vapor release of the silicon torsional resonator from the underlying sacrificial SiO$_2$ layer.   A completed device is shown in Fig.~\ref{fig1}a. 

The chip of nanomagnetic optomechanical resonators is then aligned on top of a printed circuit board patterned with drive coils \cite{Los15,Kim17} inside of a custom measurement vacuum chamber.  The drive coils allow AC excitation in both the $x$ and $z$-axes, Fig.~\ref{fig1}a, for the purpose of resonant drive for measurement of magnetization (using the $z$-axis coil) and down-mixing of high frequency spin modes (using both the $x$ and $z$-axis coils).  Optomechanical measurement is performed using dimpled tapered fiber coupling\cite{Michael07,Hau14} to the whispering gallery mode optical resonator.  Thermomechanical calibration\cite{Hauer13} reveals a torque sensitivity of 0.5 zN$\cdot$m/$\sqrt\textrm{Hz}$ sensitivity for both devices presented here, approximately four orders of magnitude better than devices used in previous TMRS measurements.\cite{Los15}  This level of torque sensitivity enables us to drive the AC excitation coils without the need for high-power amplifiers that in past measurements have limited the drive frequency,\cite{Los15} extending our measurement bandwidth above 1 GHz.  Finally, external magnetic fields are applied using a permanent magnet affixed to a long-range-travel positioning stage, and measured with a three axis Hall probe.  TMRS data is acquired by simultaneously sweeping the frequencies of UHF signals applied to $x$ and $z$-axis coils, keeping their frequency difference locked to that of the torsional resonance.  The two signal generators are synced to an external rubidium frequency standard.   The optomechanically-detected, down-mixed signal is recorded using a 50 MHz digital lock-in.  

\begin{figure}[t]
\centerline{\includegraphics[width=3.3in]{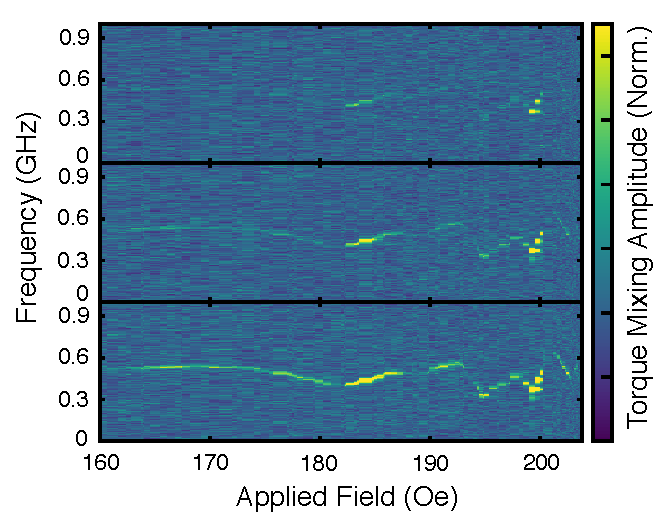}}
\caption{{\label{fig2}} Spin resonances corresponding to the gyrotropic motion of a single magnetic vortex in a 0.85 $\mu$m diameter permalloy disk at three different UHF drive powers, 3, 5, and 7 dBm from top to bottom.  The same power was applied in both the $x$ and $z$ axes.  At these levels the UHF drive fields do not alter the gyrotropic mode frequencies, revealing the unadulterated behavior.  Higher powers cause drifts in the optomechanical read-out scheme by thermally shifting the optical resonance.}
\end{figure}

It is important to consider the geometry of the applied magnetic fields and the spin texture of the magnetic structure in this experiment to understand the spin resonance that we observe.  To first order, the spins in the magnetically-soft permalloy can be considered to point in any direction.  Yet in mesoscale structures the magnetization is often dictated the boundary conditions of the disk, i.e.~shape anisotropy.  In a thin disk, as patterned here, the magnetization lies in the plane -- and curls around the boundaries of -- the disk, forming a vortex at the center of the disk in zero magnetic field.\cite{Wac02}  When a DC magnetic field is applied in the plane of the disk, this vortex is displaced orthogonal to the applied field.  For example, when a DC field is applied along the $x$-axis, the vortex is displaced along the $y$-axis, increasing the magnetization in the $x$ direction and hence the direct torque measured.  An additional AC drive field applied along the $x$-axis will drive the gyration the vortex in the plane of the disk.  This gyration can then be detected by down-mixing to the mechanical resonance frequency with an additional AC drive along the $z$-axis.

In Fig.~2, we show a sample of such a down-mixed spin resonance for a 0.85 $\mu$m diameter permalloy disk, with a thickness to radius ratio of 0.12.  In the absence of pinning, the in-plane gyration of the vortex, called the gyrotropic mode, is expected to be at 480 MHz,\cite{Novosad05} in good agreement with the average value in the current experiment.  Yet there is significant variation of the gyrotropic mode frequency with the applied in-plane magnetic field.  A variety of features can be seen, such as a range of applied fields from 160 to 180 Oe where the resonance changes slowly and smoothly, and regions -- near 200 Oe for example -- where the resonance frequency varies more significantly and with abrupt changes.  This can be understood as the vortex core, which is roughly 10 nm in diameter,\cite{Wac02} moving through defect-free regions where the resonance remains constant or regions where the vortex interacts with defects.\cite{Park03,Badea16}  Interaction with a defect, such as a variation in the film thickness,\cite{Chen12} is known to confine and pin vortices.\cite{Uhlig05,Chen12b} The gyrotropic spin mode can be thought of as a harmonic oscillation of the vortex in a potential well defined by the stiffness of the in-plane magnetization.\cite{Zar13}  When the vortex interacts with a defect it becomes more tightly localized and the stiffness increases, increasing the frequency of the gyrotropic spin mode.\cite{Compton06,Chen12b}

\begin{figure}[t]
\centerline{\includegraphics[width=3.3in]{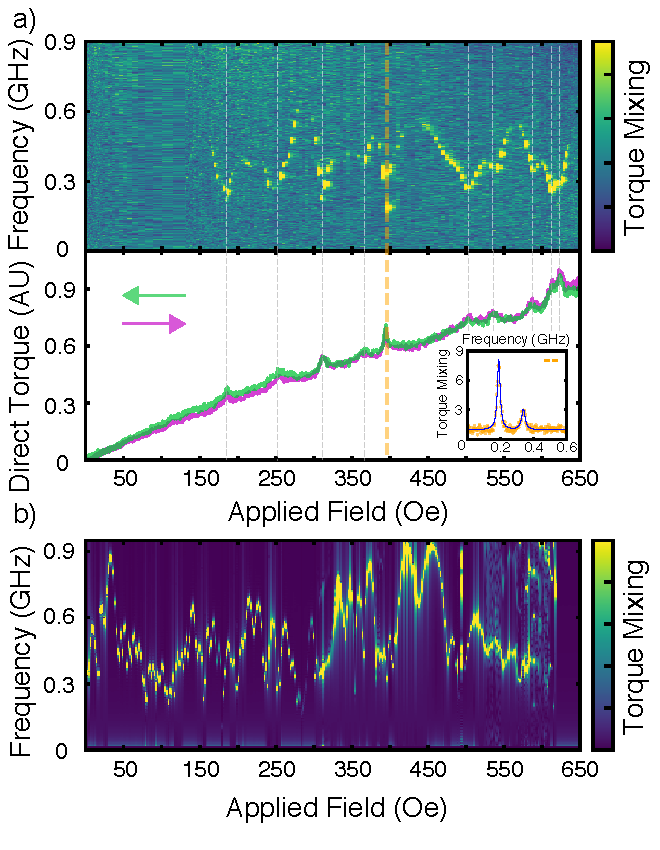}}
\caption{{\label{fig3}} a) Gyrotropic resonance (taken while sweeping from low to high in-plane field) and direct torque measurements of the in-plane magnetization and susceptibility of a 1.1 $\mu$m diameter permalloy disk.  The peaks in the direct torque correspond to increases in the susceptibility, $\partial m/\partial H$.  Indicated by the dashed lines, these points of softer magnetization also correspond to dips in the gyrotropic mode frequency, indicating that the vortex experiences a lower restoring force to in-plane motion.  Inset shows that at certain applied fields, particularly apparent when the vortex experiences a lower restoring force and gyrates at larger amplitude, multiple resonances can be observed.  Since the vortex motion is being driven by the UHF fields, these resonances can be independently accessed as the drive frequencies are swept, and correspond to the weak interaction of the vortex with multiple defects. b) Micromagnetic simulation showing qualitatively similar behavior of the gyrotropic resonance by including 20 nm polycrystalline grains with $\pm10$\% variation in the saturation magnetization. }
\end{figure}

The UHF magnetic field applied along the $x$-axis serves to drive the vortex gyrotropic processions and, in principle, could alter the potential landscape seen by the vortex\cite{Badea15} and hence the gyrotropic mode frequencies.  To test whether or not our drive fields affect the spin resonance frequencies we scanned the identical vortex pathway through the permalloy disk at multiple UHF drive powers.  Fig.~2 demonstrates that we are indeed in the regime where one can safely ignore the influence of the drive amplitude.  Residing in this regime allows us the opportunity to observe the effects of pinning, whereas at higher gyration amplitudes the vortex effectively sees a pristine landscape.\cite{Novosad05,Chen12b}  Currently higher drive powers result in heating of the printed circuit board containing the drive coils, and hence the silicon devices, causing drifts in the optomechanical detection scheme.  Future experiments mitigating this heating effect could test magnetically assisted vortex de-pinning and the effect on the gyrotropic mode frequencies.\cite{Chen12b,Badea15}

We now turn our attention to a second device, fabricated to house a 1.1 $\mu$m diameter permalloy disk.  As seen in Fig.~3 the average gyrotropic mode frequency in this disk is somewhat lower, $\sim$400 MHz, consistent with expectations from geometric scaling arguments.  A larger disk has lower magnetic stiffness since the vacuum boundaries are further away and hence lower gyrotropic mode frequency, predicted to be $\sim$385 MHz for a thickness to radius ratio of 0.09.\cite{Novosad05}  Here we additionally perform measurements of the total magnetization of the nanomagnetic disk by directly driving the torsional mode on resonance with an AC magnetic field along the $z$-axis and measuring the direct torque, Fig.~3.\cite{Davis10,Burgess13}  Furthermore, with a small component of the AC magnetic field applied along the $x$-axis, the vortex is dithered in-plane and the direct torque also encodes the in-plane magnetic susceptibility, $\partial m/\partial H$, as recently discovered in Ref.~\citenum{Wu17}.  The result is a peak in the direct torque whenever there is a change in the susceptibility, as can be seen in Fig.~3.  These susceptibility peaks can be compared with the behavior of the gyrotropic mode frequencies.  Specifically, one can see that at the applied DC magnetic fields where there is an increased susceptibility the gyrotropic mode frequency dips.  This is consistent with the picture of the vortex oscillating in a potential defined by the magnetic stiffness, as quantified by the susceptibility.  When the susceptibility increases it is easier to alter the magnetization, hence the potential well is shallower and the mode frequency lower.  

In light of this, one realizes that the magnetic fields in which the gyrotropic mode frequency increases correspond to the tightest confinement of the vortex, and likely correspond to defects in the permalloy disk.  In order to identify a possible source of these defects we compare with micromagnetic simulations (performed using mumax).\cite{Vansteenkiste14}  Fig.~3b shows simulations of the gyrotropic mode spectrum including a polycrystalline grain size of 20 nm with variations in the saturation magnetization of $\pm10$\% (around $M_{\rm s} = 800$ kA m$^{-1}$).\cite{Cullity1972}  Good qualitative agreement is seen between the simulation and measurement, with large variations in the spin mode frequencies as the vortex moves through the pinning associated with the simulated grains.  Therefore, the pinning defects observed in the measurements could originate from polycrystallinity.  We also note that these simulations suggest that the vortex annihilation should occur at an applied field just above which we used in these experiments.  Comparing with the radius of the disk, one can roughly calibrate the applied magnetic field in terms of vortex displacement.  One can then read off of Fig.~3a that a defect occurs roughly every 50 Oe, suggesting that our permalloy film contains polycrystalline grains of $\sim$40 nm.  

\begin{figure}[b]
\centerline{\includegraphics[width=3.0in]{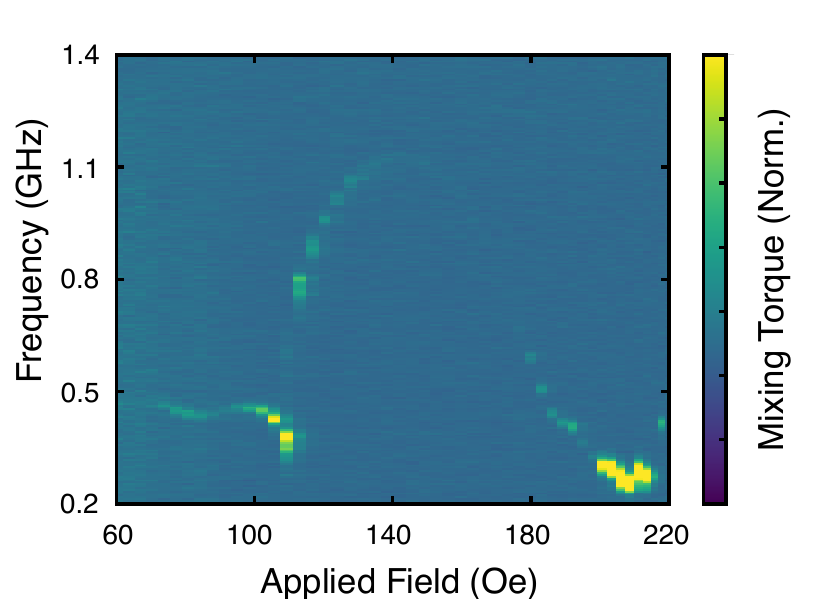}}
\caption{{\label{fig4}} A rare event in which the vortex becomes strongly pinned by a defect, driving the gyrotropic mode frequency above 1.1 GHz.  Pinning lasts over applied fields of 70 Oe, demonstrating the robustness of these pinning sites and the high frequency spin modes.  Only one pinning site of this strength was encountered in thorough exploration of the two as-grown disks.}
\end{figure}

Finally, we report on the observation of a rare event observed in the 1.1 $\mu$m diameter permalloy disk device.  At one particular orientation of DC magnetic field in the plane of the disk, we observed a strong pinning of the vortex that drove the gyrotropic mode up to $\sim$1.1 GHz.  This event, shown in Fig.~4, pinned the vortex over a field range of approximately 70 Oe, demonstrating the stability of such a pinned state to perturbations in the applied field.  Together these observations lead us to speculate that by engineering defects, as has been done with localized focused-ion-beam milling,\cite{Burgess13} one could control the spin mode resonance frequencies for potential applications.  Such as discussed above, the conversion of GHz signals to telecom wavelengths.\cite{Hil12,Bochmann13}  Additionally, the ability to simultaneously study magnetization, susceptibility, and spin resonances together are of great utility in a lab-on-a-chip approach to nanomagnetism.

In conclusion, by taking advantage of the sensitivity and bandwidth of cavity optomechanical detection, we are able to observe ultra-high frequency spin modes in permalloy disks through the technique of torque mixing resonance spectroscopy.  Combined with direct torque measurements of the in-plane susceptibility and micromagnetic simulations, we are able to show how the gyrotropic mode of individual magnetic vortices interacts with localized defects in the permalloy.  When a vortex is pinned by such defects the spin mode frequencies can jump to as high as 1.1 GHz, a phenomenon that has previously only been observed in time-resolved Kerr microscopy.\cite{Compton10,Chen12}  Opportunities exist for novel applications in control of these spin mode frequencies through the engineering of defects.  Furthermore with the TMRS sensitivity and bandwidth presented here, it may be possible to observe non-thermal, quantum effects of vortex dynamics at low temperatures.\cite{Zar13b}

This work was supported by the University of Alberta, Faculty of Science; the Natural Sciences and Engineering Research Council, Canada (Grants Nos. RGPIN-2016-04523, DAS492947-2016, and STPGP 493807); and the Canada Foundation for Innovation.

\end{document}